\documentclass[prl,showpacs,twocolumn,amsmath,amssymb,longbibliography]{revtex4-1}
\usepackage{graphicx}
\usepackage{amsmath,amsfonts}
\usepackage{dcolumn}
\usepackage{bm}
\usepackage{slashed}
\usepackage{nicefrac}
\usepackage{epstopdf}
\usepackage{color}

\begin{document}
\title{Can our spacetime emerge from anti -- de Sitter space?}
\author{Mikhail V. Altaisky}
\affiliation{Space Research Institute RAS, Profsoyuznaya 84/32, Moscow, 117997, Russia}
\email{altaisky@rssi.ru}
\author{Robin Raj}
\affiliation{Mahatma Gandhi University,  Priyadarsini Hills, Kottayam, Kerala, 686560, India}
\email{robuka97@gmail.com}
\date{Jan 17, 2022} 
\begin{abstract}
We present a model showing that our four-dimensional spacetime with the signature 
$(+,-,-,-)$ and almost vanishing positive curvature may have originated from 
a $b$-ary tree-like branching of a single discrete entity, and the 
$AdS_5$ space related to this branching process. 
\end{abstract}
\maketitle

The standard cosmology, the $\Lambda$CDM model, is based on the Big Bang theory.
The latter explains the existence of our observable Universe by a sequential 
expansion of a hot and high-density matter from quantum gravity scales  to modern size of the Universe \cite{Peebles2020}. At quantum gravity scales the spacetime is believed to be discrete, possibly in a form of spin network \cite{Penrose1971,RS1995}. For this reason 
the importance of the question of how our continuous differentiable spacetime has 
emerged from a discrete one cannot be overestimated \cite{Penrose1971}. 

The ''reverse engineering'' of the expanding Universe driven by dark energy, cold dark matter and the ordinary matter, according to the Einstein equations 
\begin{equation}
R_{\alpha\beta}-\frac{1}{2}g_{\alpha\beta}R + \Lambda g_{\alpha\beta}=-\kappa T_{\alpha\beta}, \label{ee}
\end{equation} 
leads to the Big Bang singularity.

\begin{figure}[ht]
\centering \includegraphics[width=5cm]{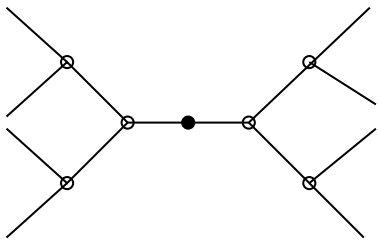}
\caption{Self-similar fragmentation of a single entity ($\bullet$) into $b$ parts according 
to a $b$-ary tree. The case of $b=2$ is shown}
\label{fg2:pic}
\end{figure}

It is possible, however, to put the problem up side down. Suppose we have a single 
entity, which brakes up into fragments according to a tree-like graph, see Fig.~\ref{fg2:pic}. 
Can we construct a four-dimensional de Sitter spacetime with the Minkowski signature by doing so for some given value of branching factor $b$? 

The $b$-ary trees, i.e., the trees with branching factor $b$, are intimately related with {\em hyperbolic spaces}. Indeed, if we define the 'circumference' of a tree as 
the number of nodes at the distance of exactly $r$ hopes from the root, this circumference is proportional to $b^r$ \cite{KPKVB2010}. At the sufficiently large $r$, this dependence coincides with the circle length in the hyperbolic space 
$
l(r) = 2\pi \sinh (\zeta r),
$
if we admit $\zeta = \ln b$.
Strictly speaking, the regularity of the tree, i.e., the exact self-similarity 
of the branching process, is not required: as soon as some classification of nodes, 
elements of a set, is at least approximately a tree the corresponding space is negatively curved \cite{Gromov2007}. This fact is widely used in studies of 
different networks in computer science, social networks, etc., see, e.g. \cite{KPKVB2010} and  references therein.

The Friedmann-Lemaitre-Robertson-Walker (FLRW) solution of the Einstein equations 
\eqref{ee} provides a homogeneous and isotropic metrics  
\begin{equation}
dl^2 = c^2 dt^2 - S^2(t) \left[
\frac{dr^2}{1-k r^2} + r^2 (d\theta^2 + \sin^2\theta d\phi^2)
\right], \label{flrw}
\end{equation}
which determines the expanding ($k\!>\!0,\Lambda\!=\!0$) Universe in terms of an ODE system 
\begin{align}\nonumber 
2\frac{\ddot{S}}{S} + \frac{\dot{S}^2+kc^2}{S^2} &= \frac{8\pi G}{c^2}T_i^i,
\quad i=\overline{1,3}\\
\frac{\dot{S}^2+kc^2}{S^2} &= \frac{8\pi G}{3c^2}T_0^0,
\end{align}
where $T_\alpha^\beta$ is the energy-momentum tensor of the matter fields, and the upper dots 
mean the derivatives with respect to the time argument $t$, as usual. 

\begin{figure}[ht]
\centering \includegraphics[height=4cm]{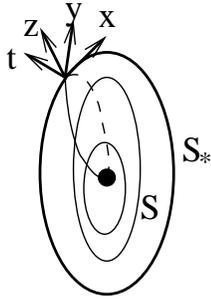}
\caption{Different paths from  'Big Bang' may result in different local time ($t$) on the same surface of constant curvature ($S_*$)}
\label{sc:pic}
\end{figure}

The $S(t)$ function is understood as an expansion rate of the Universe, experienced by any local observer at time $t$ anywhere in space. However, beyond the Eq.\eqref{flrw}-model it is not self-evident that the expansion rate should be a 
unique function of local time. For instance, the Universe may be connected, 
but not simply-connected \cite{RL1995}, and the evaluation of $S$ along different 
paths may give different results, see Fig.~\ref{sc:pic}. 
For this reason, it just seems more general to consider the radius (or the curvature) of the Universe as {\em an independent coordinate}, which traces the global history of the Universe. 

In this sense, we may live on a four-dimensional boundary of a five-dimensional gravity, which arises, for instance, in the low-dimensional limit of heteroic string theory \cite{GW1986,Bento1996}. If the Universe expansion is ceased (gedanken experiment!), our local spacetime should be Minkowskian for we observe the speed 
of light to be the same in any inertial frame of reference.

Our goal is to explain the positively curved local spacetime with the signature 
$(+,-,-,-)$ in terms of some global branching process. The situation resembles the 
inflation of a balloon, the surface of which is our spacetime. The changes 
in the matter fields, living on the balloon's surface, take place due to: ($i$)
local time evolution, taking place in the local  frame of reference; ($ii$) branching processes, which result in the balloon expansion. Thus the number of 'hops' from 
the beginning of inflation plays the role similar to the local time, but not identical to it. The measurements of local time intervals are accessible to the local observers on the balloon's surface, but the number of hops, i.e., the {\em bulk geometry}, is not accessible, unless somebody stops the inflation by ceasing the 
branching process at a given number of hops from the root. In such a case our 
toy Universe will be frozen at a surface of constant curvature. 

Since the above described dynamical structure is a tree-like, it should be described by a negatively curved five-dimensional space -- we need an extra 'time' coordinate $\rho$ to describe the branching. Such five-dimensional space can be constructed by the 
embedding of the two-time analogue of the Minkowski sphere 
\begin{equation}
x_0^2 + x_6^2 - \sum_{i=1}^4 x_i^2 = L^2 \label{a5def}
\end{equation}
into 6d Minkowski space with the metrics $$dl^2 =dx_0^2 + dx_6^2 - \sum_{i=1}^4 dx_i^2.$$
The resulting space is known as $AdS_5$ and is widely known for  $AdS/CFT$ correspondence 
in quantum field theory \cite{Maldacena1998,RT2006p}.
The $AdS_5$ space is invariant under $SO(2,4)$ symmetry group, which provides 
a natural choice of coordinates on it:
\begin{align}\nonumber 
x_0 &= L \cosh \rho \cos\tau, &\\
x_6 &= L \cosh \rho \sin\tau, & \label{x6} \\
\nonumber x_i &= L \sinh \rho\ \Omega_i, & i = \overline{1,4},
\end{align}
where $\Omega_i$ are the coordinates on the unit sphere in four dimensions. 
The hyperbolic coordinate $\rho$ is the branching coordinate we are looking 
for. So, we expect the slice $\rho=const$ to be our four-dimensional spacetime frozen 
at a given branching step. 

Casting the metrics on $AdS_5$ in the coordinates \eqref{x6}, and assuming 
$L=1$ hereafter, we get:
\begin{align}
dl^2 &= \cosh^2 \rho d\tau^2 - d\rho^2 - \sinh^2\rho d\Omega_3^2, 
\label{g5}\\
\nonumber d\Omega_3^2&= d\chi^2 + \sin^2\chi d\theta^2 + \sin^2 \chi \sin^2\theta d\phi^2 ,
\end{align}
where $d\Omega_3^2$ is the squared interval on the the unit sphere $S^3$:
\begin{align*}
x_4 &=& \cos\chi &,& x_1 &=&\sin\chi \cos\theta, \\
x_2 &=& \sin\chi \sin\theta \cos\phi&,& x_3 &=&\sin\chi \sin\theta\sin\phi.
\end{align*}

Using the coordinates $[\tau,\chi,\theta,\phi]$ with the metrics \eqref{g5} on the ($\rho=const$)-hypersurface  
we derive the local metrics of the four-dimensional universe: 
\begin{widetext}
\begin{align} 
g_{\mu\nu} &= \begin{pmatrix} 
\cosh^2\rho & 0 & 0 & 0\cr 
0 & -\sinh^2\rho & 0 & 0\cr 
0 & 0 & -\sinh^2\rho \sin^2\chi& 0 \cr 
0 & 0 & 0 & -\sinh^2\rho \sin^2\chi \sin^2\theta
\end{pmatrix}, \quad
\mu,\nu \in [\tau,\chi,\theta,\phi]. \label{g4}
\end{align}
\end{widetext}
The Christoffel symbols for the metrics \eqref{g4} are:
\begin{align*}
\Gamma^\chi_{\theta\theta} &= -\cos\chi \sin\chi, &\Gamma^\chi_{\phi\phi} &= -\cos\chi \sin\chi \sin^2\theta, \\
\Gamma^\theta_{\chi\theta}&= \Gamma^\phi_{\chi\phi} = \cot \chi, &
 \Gamma^\theta_{\phi\phi} &= -\cos\theta \sin\theta,  \Gamma^\phi_{\theta\phi} = \cot \theta.
\end{align*}
Metrics \eqref{g4} defines a positively-curved spacetime with the non-zero components 
of the Riemann tensor being 
\begin{align*} \nonumber 
R_{\phantom{3}\chi\theta\chi}^{\theta} &=& R_{\phantom{4}\chi\phi\chi}^{\phi} &=& 1, \\
R_{\phantom{2}\theta\theta\chi}^{\chi} &=&-R_{\phantom{4}\theta\phi\theta}^{\phi} &=& -\sin^2\chi \\
\nonumber
R_{\phantom{2}\phi\phi\chi}^{\chi} &=& R_{\phantom{3}\phi\phi\theta}^{\theta} &=& -\sin^2\chi \sin^2\theta,
\end{align*}
where
$$
R_{\phantom{l}ijk}^l =  \Gamma_{ik,j}^l - \Gamma_{ij,k}^l 
+ \Gamma_{ik}^m \Gamma_{mj}^l  - \Gamma_{ij}^m \Gamma_{mk}^l. 
$$
The scalar curvature $R = g^{ik}R_{ik}$ is positive for the 
metrics \eqref{g4}: 
\begin{equation}
R = \frac{6}{\sinh^2\rho}, \label{cR}
\end{equation}
with the non-zero components of the Ricci tensor $$R_{ql}= R^i_{\phantom{i}qli}$$ being
\begin{align*}
R_{\chi\chi}=-2,\ R_{\theta\theta} = -2\sin^2\chi,\ R_{\phi\phi}=-2\sin^2\chi \sin^2\theta .
\end{align*}
The non-zero components of the Einstein's tensor 
$$G_{ql}=R_{ql}-\frac{1}{2}g_{ql}R$$ 
have the form
\begin{equation}
G_\tau^\tau=-\frac{3}{\sinh^2\rho},\quad G_\chi^\chi=G_\theta^\theta=G_\phi^\phi=-\frac{1}{\sinh^2\rho}.
\label{et4}
\end{equation}

In the limit of $\rho \to \infty$ the curvature \eqref{cR} vanishes and we yield 
a flat Minkowski space, free of matter fields. 
For any finite value of $\rho$ the Einstein tensor \eqref{et4} implies the expansion of the Universe due to the initial branching process. From 
the dimensional consideration we infer  that for an $AdS_n$ space the branching 
factor is $b\!=\!n$, and hence $b\!=\!5$ in our case. 

We do not pretend to a strict proof, but if we make Euclidean change $\tau=\imath t$ in \eqref{g5}, we can map the ($\rho=const$)-hypersurface onto $S^4$ sphere. The  $S^4$ sphere admits a tessellation with four-dimensional simplices, each of those has five vertices and five hyperfaces, and can be subdivided into five parts according to the branches of a ($b\!=\!5$)-tree. This corresponds to the ($b\!=\!5$)-tree fragmentation.

The charm of the $AdS/CFT$-correspondence is due to the fact that quantum theory gravity can be described using conformal field theory on the boundary, which is a 
flat space. Classically, conformal invariance can be understood as a scaling 
invariance: the fields behave identically at different scales, as it happens for fractals, generated by a tree-like fragmentation process \cite{Feder}. In quantum field theory the 
conformal invariance is possible for massless fields. Thus the gravity is 
a kind of memory of such self-similar fragmentation, a kind of hologram of our 
world on the boundary \cite{Witten1998,Singh2018}. 

Our approach extends this point of view. Instead of considering a flat boundary 
only, we consider a collection of different slices of the discrete fragmentation process, labelled by the number of hops ($r$) from the root of the branching tree, or the hyperbolic coordinate $\rho$ in the continuous approximation. Each slice represents our four-dimensional universe at a frozen instant of inflation. To understand 
the dynamics of the whole Universe we need not only the dynamics on current slice 
with respect to the local time $\tau$, but also the history of the fragmentation 
process, i.e. which regions of the previous slices were the 'parents' of the current slice regions. This is deeply related to the entanglement entropy \cite{RT2006}, and 
can be considered only within the framework of quantum field theory.
In our approach 
the 'static' solution $\rho=const$ does make sense, since we can fill our four-dimensional  space-time with quantum fields $\phi(t,x,y,z)$ and consider their 
dynamics with respect to the local time $t$.

We do not neglect the standard FLRW solution here, instead, using the Weyl postulate on the trajectories from the Big Bang, we can study the case 
of 
\begin{equation}
dS = \frac{\partial S}{\partial \tau} d\tau, \label{W}
\end{equation}
where 
\begin{equation}
S = \sinh \rho,
\end{equation}
according to the metrics \eqref{g5}. The equality \eqref{W} in our context is 
rather strong physical assertion. It says that any change in the local time $\tau$, 
and hence any evolution of local physical fields, happens {\em synchronously and due to} the changes in $\rho$ -- the Universe expansion in our case. We are not 
going to dig into philosophical problems related to the assumption \eqref{W}, instead we will study some formal consequences of it in a five-dimensional world 
with the metrics \eqref{g5}.

Since the restriction \eqref{W}  
makes the metrics \eqref{g5} into four-dimensional metrics 
\begin{align}\nonumber 
dl^2 &= \left[1+S^2 - \frac{\dot{S}^2}{1+S^2} \right] d\tau^2 \\
     &- S^2 \left[ 
      \frac{dr^2}{1-r^2} + r^2 (d\theta^2 + \sin^2\theta d\phi^2)\right],
\label{g4w}
\end{align}
where $r=\sin\chi$, and $d\rho = \frac{dS}{\sqrt{1+S^2}}$, up to conformal 
transformation, the metrics \eqref{g4w} is equivalent 
to the standard FLRW metrics \eqref{flrw}. Its classical dynamics in the original time $\tau$ can be studied using the Einstein tensor 
\begin{align}\nonumber
G_i^i &= \frac{
(S^2+1)^3 \dot{S}^2 + 3S^2 \dot{S}^4 -2S(S^2+1)^3 \ddot{S} -(S^2+1)^4
}{
S^2 \left[\dot{S}^2 -(S^2+1)^2 \right]^2
}, \\
G_0^0 &= \frac{3}{S^2} \cdot \frac{(\dot{S}S)^2 + (S^2+1)^2}{\dot{S}^2 - (S^2+1)^2},\quad i=\overline{1,3},
\end{align} 
which follows from the metrics \eqref{g4w}. 
The analysis of the classical dynamics may be the subject of a separate study,
but is beyond the scope of this paper.

Planning the quantum field theory studies for future research, we can still say 
within presented classical framework that the Friedmann model description in terms of the four-dimensional 
spacetime coordinated is too small, and we may need a new {\em independent} inflation coordinate $\rho$ to describe our expanding Universe. 

Having this note almost prepared the authors became aware of a new paper 
\cite{BDH2021}, proposing an exponential tree-like growth of the dark matter ($\chi$) number 
density by using the energy of the heat bath particles ($\psi$): $\chi \psi \to \chi\chi$. In some sense our paper presents an alternative to this approach to 
the dark matter abundance: the exponential growth may be attributed to the 
evolution of spacetime from a discrete to a continuous one. The matter fields then should be added to this background.

\paragraph*{Acknowledgement} The authors are thankful to Dr. O.Tsupko, Prof. C.Rovelli and Prof. E.Shchepin for useful discussions, and to Prof. I.Oseledets for useful references.
\paragraph*{Statements on compliance with ethical standards}
Both authors were paid from budget with no external funding to be acknowledged. The authors declare that they have no conflicts of interests. 

\begin{thebibliography}{15}%
\makeatletter
\providecommand \@ifxundefined [1]{%
 \@ifx{#1\undefined}
}%
\providecommand \@ifnum [1]{%
 \ifnum #1\expandafter \@firstoftwo
 \else \expandafter \@secondoftwo
 \fi
}%
\providecommand \@ifx [1]{%
 \ifx #1\expandafter \@firstoftwo
 \else \expandafter \@secondoftwo
 \fi
}%
\providecommand \natexlab [1]{#1}%
\providecommand \enquote  [1]{``#1''}%
\providecommand \bibnamefont  [1]{#1}%
\providecommand \bibfnamefont [1]{#1}%
\providecommand \citenamefont [1]{#1}%
\providecommand \href@noop [0]{\@secondoftwo}%
\providecommand \href [0]{\begingroup \@sanitize@url \@href}%
\providecommand \@href[1]{\@@startlink{#1}\@@href}%
\providecommand \@@href[1]{\endgroup#1\@@endlink}%
\providecommand \@sanitize@url [0]{\catcode `\\12\catcode `\$12\catcode
  `\&12\catcode `\#12\catcode `\^12\catcode `\_12\catcode `\%12\relax}%
\providecommand \@@startlink[1]{}%
\providecommand \@@endlink[0]{}%
\providecommand \url  [0]{\begingroup\@sanitize@url \@url }%
\providecommand \@url [1]{\endgroup\@href {#1}{\urlprefix }}%
\providecommand \urlprefix  [0]{URL }%
\providecommand \Eprint [0]{\href }%
\providecommand \doibase [0]{http://dx.doi.org/}%
\providecommand \selectlanguage [0]{\@gobble}%
\providecommand \bibinfo  [0]{\@secondoftwo}%
\providecommand \bibfield  [0]{\@secondoftwo}%
\providecommand \translation [1]{[#1]}%
\providecommand \BibitemOpen [0]{}%
\providecommand \bibitemStop [0]{}%
\providecommand \bibitemNoStop [0]{.\EOS\space}%
\providecommand \EOS [0]{\spacefactor3000\relax}%
\providecommand \BibitemShut  [1]{\csname bibitem#1\endcsname}%
\let\auto@bib@innerbib\@empty
\bibitem [{\citenamefont {Peebles}(2020)}]{Peebles2020}%
  \BibitemOpen
  \bibfield  {author} {\bibinfo {author} {\bibfnamefont {P.J.E.}\ \bibnamefont
  {Peebles}},\ }\href@noop {} {\emph {\bibinfo {title} {Principles of Physical
  Cosmology}}}\ (\bibinfo  {publisher} {Princeton University Press},\ \bibinfo
  {year} {2020})\BibitemShut {NoStop}%
\bibitem [{\citenamefont {Penrose}(1971)}]{Penrose1971}%
  \BibitemOpen
  \bibfield  {author} {\bibinfo {author} {\bibfnamefont {R.}~\bibnamefont
  {Penrose}},\ }\bibfield  {title} {\enquote {\bibinfo {title} {Angular
  momentum: an approach to combinatorial spacetime},}\ }in\ \href@noop {}
  {\emph {\bibinfo {booktitle} {Quantum Theory and Beyond}}},\ \bibinfo
  {editor} {edited by\ \bibinfo {editor} {\bibfnamefont {T.}~\bibnamefont
  {Bastin}}}\ (\bibinfo  {publisher} {Cambridge University Press},\ \bibinfo
  {address} {Cambridge, England},\ \bibinfo {year} {1971})\ pp.\ \bibinfo
  {pages} {151--180}\BibitemShut {NoStop}%
\bibitem [{\citenamefont {Rovelli}\ and\ \citenamefont
  {Smolin}(1995)}]{RS1995}%
  \BibitemOpen
  \bibfield  {author} {\bibinfo {author} {\bibfnamefont {C.}~\bibnamefont
  {Rovelli}}\ and\ \bibinfo {author} {\bibfnamefont {L.}~\bibnamefont
  {Smolin}},\ }\bibfield  {title} {\enquote {\bibinfo {title} {Spin networks
  and quantum gravity},}\ }\href {\doibase 10.1103/PhysRevD.52.5743} {\bibfield
   {journal} {\bibinfo  {journal} {Phys. Rev. D}\ }\textbf {\bibinfo {volume}
  {52}},\ \bibinfo {pages} {5743--5759} (\bibinfo {year} {1995})}\BibitemShut
  {NoStop}%
\bibitem [{\citenamefont {Krioukov}\ \emph {et~al.}(2010)\citenamefont
  {Krioukov}, \citenamefont {Papadopoulos}, \citenamefont {Kitsak},
  \citenamefont {Vahdat},\ and\ \citenamefont {Bogu\~n\'a}}]{KPKVB2010}%
  \BibitemOpen
  \bibfield  {author} {\bibinfo {author} {\bibfnamefont {D.}~\bibnamefont
  {Krioukov}}, \bibinfo {author} {\bibfnamefont {F.}~\bibnamefont
  {Papadopoulos}}, \bibinfo {author} {\bibfnamefont {M.}~\bibnamefont
  {Kitsak}}, \bibinfo {author} {\bibfnamefont {A.}~\bibnamefont {Vahdat}}, \
  and\ \bibinfo {author} {\bibfnamefont {M.}~\bibnamefont {Bogu\~n\'a}},\
  }\bibfield  {title} {\enquote {\bibinfo {title} {Hyperbolic geometry of
  complex networks},}\ }\href {\doibase 10.1103/PhysRevE.82.036106} {\bibfield
  {journal} {\bibinfo  {journal} {Phys. Rev. E}\ }\textbf {\bibinfo {volume}
  {82}},\ \bibinfo {pages} {036106} (\bibinfo {year} {2010})}\BibitemShut
  {NoStop}%
\bibitem [{\citenamefont {Gromov}(2007)}]{Gromov2007}%
  \BibitemOpen
  \bibfield  {author} {\bibinfo {author} {\bibfnamefont {M.}~\bibnamefont
  {Gromov}},\ }\href@noop {} {\emph {\bibinfo {title} {Metric structures for
  {R}iemannian and Non-{R}iemanian spaces}}}\ (\bibinfo  {publisher}
  {Birkh\"auser},\ \bibinfo {address} {Boston},\ \bibinfo {year}
  {2007})\BibitemShut {NoStop}%
\bibitem [{\citenamefont {Lach\'ieze-Rey}\ and\ \citenamefont
  {Luminet}(1995)}]{RL1995}%
  \BibitemOpen
  \bibfield  {author} {\bibinfo {author} {\bibfnamefont {M.}~\bibnamefont
  {Lach\'ieze-Rey}}\ and\ \bibinfo {author} {\bibfnamefont {J.-P.}\
  \bibnamefont {Luminet}},\ }\bibfield  {title} {\enquote {\bibinfo {title}
  {Cosmic topology},}\ }\href@noop {} {\bibfield  {journal} {\bibinfo
  {journal} {Phys. Rep.}\ }\textbf {\bibinfo {volume} {254}},\ \bibinfo {pages}
  {135--214} (\bibinfo {year} {1995})}\BibitemShut {NoStop}%
\bibitem [{\citenamefont {Gross}\ and\ \citenamefont {Witten}(1986)}]{GW1986}%
  \BibitemOpen
  \bibfield  {author} {\bibinfo {author} {\bibfnamefont {D.J.}\ \bibnamefont
  {Gross}}\ and\ \bibinfo {author} {\bibfnamefont {E.}~\bibnamefont {Witten}},\
  }\bibfield  {title} {\enquote {\bibinfo {title} {Superstring modifications of
  einstein's equations},}\ }\href@noop {} {\bibfield  {journal} {\bibinfo
  {journal} {Nucl. Phys. B}\ }\textbf {\bibinfo {volume} {277}},\ \bibinfo
  {pages} {1} (\bibinfo {year} {1986})}\BibitemShut {NoStop}%
\bibitem [{\citenamefont {Bento}\ and\ \citenamefont
  {Bertolami}(1996)}]{Bento1996}%
  \BibitemOpen
  \bibfield  {author} {\bibinfo {author} {\bibfnamefont {M.C.}\ \bibnamefont
  {Bento}}\ and\ \bibinfo {author} {\bibfnamefont {O.}~\bibnamefont
  {Bertolami}},\ }\bibfield  {title} {\enquote {\bibinfo {title} {Maximally
  symmetric cosmological solutions of higher-curvature string effective
  theories with dilatons},}\ }\href@noop {} {\bibfield  {journal} {\bibinfo
  {journal} {Phys. Lett. B}\ }\textbf {\bibinfo {volume} {368}},\ \bibinfo
  {pages} {198--201} (\bibinfo {year} {1996})}\BibitemShut {NoStop}%
\bibitem [{\citenamefont {Maldacena}(1998)}]{Maldacena1998}%
  \BibitemOpen
  \bibfield  {author} {\bibinfo {author} {\bibfnamefont {J.}~\bibnamefont
  {Maldacena}},\ }\bibfield  {title} {\enquote {\bibinfo {title} {The large $n$
  limit of superconformal field theories and supergravity},}\ }\href@noop {}
  {\bibfield  {journal} {\bibinfo  {journal} {Adv. Theor. Math. Phys.}\
  }\textbf {\bibinfo {volume} {2}},\ \bibinfo {pages} {231} (\bibinfo {year}
  {1998})}\BibitemShut {NoStop}%
\bibitem [{\citenamefont {Ryu}\ and\ \citenamefont
  {Takayanagi}(2006{\natexlab{a}})}]{RT2006p}%
  \BibitemOpen
  \bibfield  {author} {\bibinfo {author} {\bibfnamefont {S.}~\bibnamefont
  {Ryu}}\ and\ \bibinfo {author} {\bibfnamefont {T.}~\bibnamefont
  {Takayanagi}},\ }\bibfield  {title} {\enquote {\bibinfo {title} {Holographic
  derivation of entanglement entropy from the anti--de sitter space/conformal
  field theory correspondence},}\ }\href {\doibase
  10.1103/PhysRevLett.96.181602} {\bibfield  {journal} {\bibinfo  {journal}
  {Phys. Rev. Lett.}\ }\textbf {\bibinfo {volume} {96}},\ \bibinfo {pages}
  {181602} (\bibinfo {year} {2006}{\natexlab{a}})}\BibitemShut {NoStop}%
\bibitem [{\citenamefont {Feder}(1988)}]{Feder}%
  \BibitemOpen
  \bibfield  {author} {\bibinfo {author} {\bibfnamefont {J.}~\bibnamefont
  {Feder}},\ }\href@noop {} {\emph {\bibinfo {title} {Fractals}}}\ (\bibinfo
  {publisher} {Plenum Press},\ \bibinfo {year} {1988})\BibitemShut {NoStop}%
\bibitem [{\citenamefont {Witten}(1998)}]{Witten1998}%
  \BibitemOpen
  \bibfield  {author} {\bibinfo {author} {\bibfnamefont {E.}~\bibnamefont
  {Witten}},\ }\bibfield  {title} {\enquote {\bibinfo {title} {Anti de {S}itter
  space and holography},}\ }\href@noop {} {\bibfield  {journal} {\bibinfo
  {journal} {Adv. Theor. Math. Phys.}\ }\textbf {\bibinfo {volume} {2}},\
  \bibinfo {pages} {253--291} (\bibinfo {year} {1998})}\BibitemShut {NoStop}%
\bibitem [{\citenamefont {Singh}(2018)}]{Singh2018}%
  \BibitemOpen
  \bibfield  {author} {\bibinfo {author} {\bibfnamefont {S.}~\bibnamefont
  {Singh}},\ }\bibfield  {title} {\enquote {\bibinfo {title} {Tensor network
  state correspondence and holography},}\ }\href {\doibase
  10.1103/PhysRevD.97.026012} {\bibfield  {journal} {\bibinfo  {journal} {Phys.
  Rev. D}\ }\textbf {\bibinfo {volume} {97}},\ \bibinfo {pages} {026012}
  (\bibinfo {year} {2018})}\BibitemShut {NoStop}%
\bibitem [{\citenamefont {Ryu}\ and\ \citenamefont
  {Takayanagi}(2006{\natexlab{b}})}]{RT2006}%
  \BibitemOpen
  \bibfield  {author} {\bibinfo {author} {\bibfnamefont {S.}~\bibnamefont
  {Ryu}}\ and\ \bibinfo {author} {\bibfnamefont {T.}~\bibnamefont
  {Takayanagi}},\ }\bibfield  {title} {\enquote {\bibinfo {title} {Aspects of
  holographic entanglement entropy},}\ }\href {\doibase
  10.1088/1126-6708/2006/08/045} {\bibfield  {journal} {\bibinfo  {journal}
  {Journal of High Energy Physics}\ }\textbf {\bibinfo {volume} {2006}},\
  \bibinfo {pages} {045--045} (\bibinfo {year}
  {2006}{\natexlab{b}})}\BibitemShut {NoStop}%
\bibitem [{\citenamefont {Bringmann}\ \emph {et~al.}(2021)\citenamefont
  {Bringmann}, \citenamefont {Depta}, \citenamefont {Hufnagel}, \citenamefont
  {Ruderman},\ and\ \citenamefont {Schmidt-Hoberg}}]{BDH2021}%
  \BibitemOpen
  \bibfield  {author} {\bibinfo {author} {\bibfnamefont {T.}~\bibnamefont
  {Bringmann}}, \bibinfo {author} {\bibfnamefont {P.~F.}\ \bibnamefont
  {Depta}}, \bibinfo {author} {\bibfnamefont {M.}~\bibnamefont {Hufnagel}},
  \bibinfo {author} {\bibfnamefont {J.~T.}\ \bibnamefont {Ruderman}}, \ and\
  \bibinfo {author} {\bibfnamefont {K.}~\bibnamefont {Schmidt-Hoberg}},\
  }\bibfield  {title} {\enquote {\bibinfo {title} {Dark matter from exponential
  growth},}\ }\href {\doibase 10.1103/PhysRevLett.127.191802} {\bibfield
  {journal} {\bibinfo  {journal} {Phys. Rev. Lett.}\ }\textbf {\bibinfo
  {volume} {127}},\ \bibinfo {pages} {191802} (\bibinfo {year}
  {2021})}\BibitemShut {NoStop}%
\end{thebibliography}
%
\end{document}